\documentclass[12pt]{article}

\usepackage[utf8]{inputenc}
\pdfoutput=1

\usepackage{jheppub}

\usepackage{color}
\usepackage{graphicx}   
\usepackage{bm}
\usepackage{amsmath}
\usepackage{amsfonts}
\usepackage{eufrak}

\title{Simple non-perturbative resummation schemes beyond mean-field: case study for scalar $\phi^4$ theory in 1+1 dimensions}

\author{Paul Romatschke$^{1,2}$}
\affiliation{$^1$ Department of Physics, University of Colorado, Boulder, Colorado 80309, USA}
\affiliation{$^2$ Center for Theory of Quantum Matter, University of Colorado, Boulder, Colorado 80309, USA}
\emailAdd{paul.romatschke@colorado.edu}

\abstract{I present a sequence of non-perturbative approximate solutions for scalar $\phi^4$ theory for arbitrary interaction strength, which contains, but allows to systematically improve on, the familiar mean-field approximation. This sequence of approximate solutions is apparently well-behaved and numerically simple to calculate since it only requires the evaluation of (nested) one-loop integrals. To test this resummation scheme, the case of $\phi^4$ theory in 1+1 dimensions is considered, finding approximate agreement with known results for the vacuum energy and mass gap up to the critical point. Because it can be generalized to other dimensions, fermionic fields and finite temperature, the resummation scheme could potentially become a useful tool for calculating non-perturbative properties approximately in certain quantum field theories.}

\begin{document}

\maketitle

\section{Introduction}

There exist two main avenues to analytically calculating general observables in quantum field theories. These are based on our ability to exactly solve free (non-interacting) quantum field theories and certain gauge theories in the limit of a large number of colors and infinite interaction strength using the conjectured gauge/gravity duality.

Perturbations around the exactly solvable cases of zero and infinite interaction strength may also be calculated, but strict perturbative series expansions prove to be divergent in most cases. Thus, quantitatively reliable results away from zero and infinite interaction strength either require the use of non-perturbative resummations, or a-posteriori justification through either experimental or numerical methods such as lattice field theory\footnote{Notable exceptions are quantities that are protected by symmetries or quantities that can be accessed by taking the number of components to infinity.}.

However, there are several physics cases where one would want to have access to quantitatively accurate results for the case of intermediate coupling strengths, such as transport coefficients. This begs the question of whether it could be possible to develop new, non-perturbative techniques for quantum field theories that allow the calculations of observables at arbitrary coupling strength at least \textit{approximately}.

An example of a non-perturbative resummation scheme is the so-called mean-field approximation where an infinite number of perturbative terms are resummed into an interaction-strength dependent mass term. While wildly successful, the mean-field approximation suffers from the problem that certain interesting types of observables, such as order/disorder transitions or transport coefficients, are determined by contributions which are formally higher order in perturbation theory, and hence are not captured by the mean-field approach.

The present work is meant as a step towards developing new, non-perturbative resummations schemes that could allow one to access interesting observables at intermediate couplings approximately rather than exactly, say to within 30 percent accuracy.

To keep the discussion simple, for this work I selected one theory (single component scalar theory with quartic interaction) and one particular setup in which mean-field theory fails (symmetry breaking at zero temperature in 1+1 dimensions). However, the methods presented here are straightforward to implement in other dimensions, for fermions, at finite temperature and for other observables\footnote{I know this because I have used these methods in other contexts, calculating results that will be presented elsewhere.}. 

\section{Setup}

I consider the path integral formulation of $\phi^4$ theory in two Euclidean dimensions given by
\begin{equation}
  Z=\int {\cal D}\phi e^{-S}\,,\quad S= \int d^{2}X \left(\frac{1}{2}\partial_\mu \phi \partial_\mu \phi+\frac{m^2}{2}\phi^2+\lambda \phi^4\right)\,,
\end{equation}
where it should be noted that $\lambda$ here has mass dimension $2$ and I assume $m^2>0$ here\footnote{It is straightforward to generalize the discussion to $m^2<0$ by first splitting $\phi=\phi_0+\phi^\prime$ with $\phi_0=\langle \phi \rangle$, and then proceeding as for $m^2>0$.}. The theory is formulated at finite temperature $T$ by having one of the Euclidean directions (say $x_0$) be compactified on a circle of radius $\beta\equiv T^{-1}$. (Note that only the zero temperature limit will be considered in this work.)
Introducing two auxiliary fields $\sigma,\zeta$ using $1=\int {\cal D}\sigma \delta(\sigma-\phi^2)=\int {\cal D}\sigma {\cal D}\zeta e^{i \int \zeta (\sigma-\phi^2)}$ and subsequently integrating out $\sigma$ leads to an alternative representation of the partition function as
\begin{equation}
  Z=\int {\cal D}\phi {\cal D}\zeta e^{-S}\,,\quad S= \int d^2X \left(\frac{1}{2}\partial_\mu \phi \partial_\mu \phi+\frac{m^2}{2}\phi^2+i \zeta \phi^2 + \frac{\zeta^2}{4 \lambda}\right)\,,
\end{equation}
which is well-known. In this form, it is possible to integrate out the original field $\phi$ exactly, but the resulting determinant of the operator
\begin{equation}
  \label{eq:op}
  L=\partial_\mu\partial_\mu+m^2+2 i \zeta
\end{equation}
is in practice too unwieldy to handle for the subsequent integration over $\zeta$, thus preventing an exact solution to the field theory. Except for the global zero-mode of $\zeta$.

Writing $\zeta(X)=\frac{1}{2}\zeta_0+\zeta^\prime(X)$ with $\int d^2 X \zeta(X)=\frac{\beta V \zeta_0}{2}$ where $V$ is the ``volume'' of the Euclidean direction $x_1$, one finds
\begin{equation}
  \label{eq:zgen}
  Z=\sqrt{\frac{\beta V}{16 \lambda \pi}}\int d\zeta_0 e^{-\frac{\zeta_0^2 \beta V}{16 \lambda}}\int {\cal D}\phi {\cal D}\zeta^\prime e^{-S_0-S_I}\,,
\end{equation}
where
\begin{equation}
  S_0=\frac{1}{2}\int d^2X\left[ \partial_\mu \phi \partial_\mu \phi+m^2 \phi^2+ i \zeta_0 \phi^2+\frac{\zeta^{\prime 2}}{2\lambda}\right]\,,\quad
  S_I=i \int d^2X \zeta^\prime \phi^2\,.
  \end{equation}
Note that neglecting the contribution $S_I$ in the action, the resulting approximation becomes exactly solvable because the zero-mode contribution $\zeta_0$ only enters in the diagonal of the operator $L$, thus effectively taking the role of an effective mass contribution\footnote{This means that a theory with interaction $S_I=\lambda \int d^2X d^2Y \phi^2(x)\phi^2(y)$ can be solved exactly.}. Despite being extremely simple, the resulting approximation thus retains non-perturbative character at non-zero value of the interaction strength $\lambda$, because it corresponds to a resummation of an infinite number of diagrams in the standard perturbative approach. Therefore, I consider (\ref{eq:zgen}) with $S_I=0$ the starting point of the sequence of resummation schemes discussed in this work, and will refer to it as ``Resummation Level Zero'' (``R0'') in the following.

\subsection{Resummation Level Zero (``R0'')}

Using standard methods in field theory \cite{Laine:2016hma}, evaluating the partition function in the R0 approximation is no more difficult than the corresponding tree-level calculation in standard perturbation theory. In the zero temperature limit one finds
\begin{equation}
  Z_{R0}=\sqrt{\frac{\beta V}{16 \lambda \pi}}\int d\zeta_0 e^{-\beta V \left[\frac{\zeta_0^2}{16 \lambda}+J_0(\sqrt{m^2+i \zeta_0}\right]}\,,
\end{equation}
where in $d=1-2\epsilon$ Euclidean space dimensions
\begin{equation}
  J_0(\alpha)=\frac{\mu^{2 \epsilon}\Gamma(-\frac{1+d}{2})}{2 (4\pi)^{d/2} \Gamma\left(-\frac{1}{2}\right) \alpha^{-1-d}}=\frac{\alpha^2}{8 \pi \epsilon}+\frac{\alpha^2}{8\pi} \ln \frac{\bar \mu^2 e^1}{\alpha^2} + {\cal O}(\epsilon)\,,
\end{equation}
$\bar \mu^2=4\pi \mu^2 e^{-\gamma_E}$ is the $\overline{\rm MS}$  scale parameter. In the large volume $V\rightarrow \infty$ limit, the remaining integral over $\zeta_0$ can be evaluated exactly from the saddle point of the action $i \zeta_0=z^*$ given by $z^*_{R0}=4 \lambda I_0\left(\sqrt{m^2+z^*_{R0}}\right)$, 
where $I_0(\alpha)=2 \frac{d J_0(\alpha)}{d \alpha^2}$.
%
The two-point function 
\begin{equation}
  \langle \phi(X) \phi(Y)\rangle=\sqrt{\frac{\beta V}{16 \lambda \pi}}\int d\zeta_0 e^{-\frac{\zeta_0^2 \beta V}{16 \lambda}}G(X-Y,i \zeta_0)=G(X-Y,z^*)
\end{equation}
in the R0 approximation is similarly easy to compute, and one finds \cite{Laine:2016hma}
\begin{equation}
    G_{R0}(X,z^*_{R0})=\int_K \frac{e^{i K\cdot X}}{K^2+m^2+z^*_{R0}}\,,
\end{equation}
where $K=\left(k_0,k\right)$, $K\cdot X\equiv k_0 x_0+k x_1$ and here and in the following
\begin{equation}
  \int_K \equiv \bar \mu^{2\epsilon}(4\pi)^{-\epsilon}e^{\gamma_E \epsilon}\int \frac{d^{d+1}K}{(2 \pi)^{d+1}} \,.
\end{equation}
Since the pole mass $M$ of the two-point function $G(X,z^*)$ is an observable in principle, the sum $m^2+z^*_{R0}$ must be finite and hence in the $\overline{\rm MS}$ scheme I am led to introduce the renormalized mass $m_R$ at R0 level as
\begin{equation}
m_R^2=  m^2+\frac{\lambda}{\pi \epsilon}\,,
\end{equation}
such that
\begin{equation}
m^2+z_{R0}^*=m_R^2+\frac{\lambda}{\pi}\ln \frac{\bar \mu^2}{m^2+z_{R0}^*}\,,
\end{equation}
which has the form of a non-perturbative ``gap-equation''. In order to make contact with other non-perturbative studies of $\phi^4$ in 2 dimensions I adopt the scheme choice $\bar\mu^2=m_R^2$ such that a solution of the above R0-level gap equation is $m^2+z_{R0}^*=m_R^2$.

Note that the R0 approximation coincides with the leading-order approximation for $\phi^4$ theory with a large number of components, cf. Ref.~\cite{Carmi:2018qzm}.

\subsection{Resummation Level One (``R1'') a.k.a. Mean-Field Resummation}

In order to improve on the R0-level approximation, contributions from the fluctuation field $\zeta^\prime$ in the determinant of the operator (\ref{eq:op}) need to be taken into account. Expanding $e^{-S_I}$ perturbatively in powers of $\zeta^\prime$ would be an option in principle, except that it would ruin the non-perturbative character of the approximation. An alternative approach is to re-write the theory on the level of the action by formally performing the replacement
\begin{eqnarray}
  \label{eq:r1level}
  &S_0+S_I\rightarrow S_0^\prime +S_I^\prime\,,&\\
  &S_0^\prime=\frac{1}{2}\int d^2X\left[ \partial_\mu \phi \partial_\mu \phi+m^2 \phi^2+ i \zeta_0 \phi^2+\nu^2 \phi^2+\frac{\zeta^{\prime 2}}{2\lambda}\right]\,,\quad
  S_I^\prime=\int d^2X \left[i\zeta^\prime \phi^2-\frac{\nu^2}{2}\phi^2\right]\,.\nonumber&
  \end{eqnarray}
in (\ref{eq:zgen}) with an unknown constant mass parameter $\nu$. To fix $\nu$, it is sufficient to calculate the two-point function $\langle \phi(X)\phi(Y)\rangle$ to first non-trivial order in the fluctuation field $\zeta^\prime$ and the counter-term $\nu^2$, finding
\begin{equation}
  \label{eq:r1geq}
  G(X,i \zeta_0)=G(X,i \zeta_0)-\int d^2Y G(X-Y,i\zeta_0)G(-Y,i\zeta_0)\left[
    8\lambda G(0,i\zeta_0)-\nu^2\right]\,,
\end{equation}
which in turn suggests
\begin{equation}
  \nu^2=8 \lambda G_{R1}(0,i \zeta_0)=8 \lambda I_0(\sqrt{m^2+i\zeta_0+\nu^2})\,.
  \end{equation}
Similarly to the R0 approximation discussed above, the effective mass term $\nu^2$ is determined non-perturbatively through a gap-equation, effectively resumming an additional infinite number of diagrams in the standard perturbative approach. Therefore, I consider Eq.~(\ref{eq:zgen}) with the re-organization (\ref{eq:r1level}) the first correction to R0 and will refer to it as ``Resummation Level One'' (``R1'') in the following. It is not hard to verify that the R1 approximation is nothing else but the familiar mean-field resummation.

The partition function in the R1 approximation is obtained by including contributions to first non-trivial order in the fluctuation field $\zeta^\prime$ and the counter-term $\nu^2$, finding
\begin{equation}
  Z_{R1}=\sqrt{\frac{\beta V}{16 \lambda \pi}}\int d\zeta_0 e^{-\beta V \left[\frac{\zeta_0^2}{16 \lambda}+J_0(\sqrt{m^2+\nu^2+i \zeta_0})-2 \lambda G_{R1}(0,i\zeta_0)^2\right]}\,.
  \end{equation}
As before, in the large volume limit the integral can be evaluated exactly through its saddle point $i\zeta_0=z^*$ given by
\begin{equation}
  \label{eq:r1zeq}
  z_{R1}^*=4 \lambda I_0(\sqrt{m^2+i\zeta_0+\nu^2})=4 \lambda G_{R1}(0,z_{R1}^*)\,,
\end{equation}
where it should be noted that the contribution $\frac{\partial \nu^2}{\partial i \zeta_0}$ originating from the function $J_0$ in the exponent exactly cancels against that from $2\lambda G_{R1}^2(0,i\zeta_0)=\frac{\nu^4}{32 \lambda}$. The two-point function in the R1 approximation given by
\begin{equation}
  G_{R1}(X,z^*_{R1})=\int_K \frac{e^{i K\cdot X}}{K^2+m^2+\nu^2+z_{R1}^*}\,,
\end{equation}
then gives rise to a finite pole mass $M$ in the R1 approximation if
\begin{equation}
  \label{eq:rencon}
  m_R^2=m^2+\frac{3 \lambda}{\pi \epsilon}\,,
\end{equation}
such that
\begin{equation}
  m^2+\nu^2+z_{R1}^*=m_R^2+\frac{3 \lambda}{\pi} \ln \frac{\bar \mu^2}{m^2+\nu^2+z_{R1}^*}\,.
  \end{equation}
With the additional scheme choice $\bar\mu^2=m_R^2$ this corresponds to the renormalization condition adopted in the lattice studies of $\phi^4$ in 2 dimensions, cf. Refs.~\cite{Loinaz:1997az,Schaich:2009jk,Bosetti:2015lsa}.

\subsection{Resummation Level Two (``R2'')}

Going beyond the mean-field resummation R1 requires resumming contributions originating from higher powers of the fluctuation field $\zeta^\prime$ in the determinant of the operator (\ref{eq:op}). Taking the case of the R1-level resummation as an example, a further refinement of the R1 scheme can be defined by introducing dynamical propagators for both the $\phi,\zeta^\prime$ fields through rewriting the action (\ref{eq:zgen}) as
\begin{eqnarray}
  \label{eq:r2level}
  &S_0+S_I\rightarrow S_0^{\prime\prime} +S_I^{\prime\prime}\,,&\\
  &S_0^{\prime\prime}=\frac{1}{2}\int d^2X d^2 Y\left[\phi(X) G^{-1}(X-Y,i\zeta_0)\phi(Y)+\zeta^\prime(X)D^{-1}(X-Y)\zeta^{\prime}(Y)\right]\,,&\nonumber\\
  &S_I^{\prime\prime}=i\int d^2X \zeta^\prime(X) \phi^2(X)-\frac{1}{2}\int d^2X d^2Y \left[\phi(X)\Pi(X-Y)\phi(Y)+\zeta^\prime(X)\Sigma(X-Y)\zeta^\prime(Y)\right]\,,\nonumber&\\
  &G^{-1}(X,i\zeta_0)=G_{R0}^{-1}(X,i\zeta_0)+\Pi(X)\,,\quad
  \left(-\partial_\mu\partial_\mu+m^2+i\zeta_0\right) G_{R0}(X,i\zeta_0)=\delta(X)\,,&\nonumber\\
  &D^{-1}(X)=D^{-1}_{R0}(X)+\Sigma(X)\,,\quad D^{-1}_{R0}(X)=\frac{\delta(X)}{2 \lambda}\,,
\end{eqnarray}
with unknown self-energies $\Pi(X),\Sigma(X)$. To fix the self-energies, and thus the two-point functions $G(X),D(X)$, I calculate the $\langle \phi(X)\phi(Y)\rangle$ and $\langle \zeta^\prime(X)\zeta^\prime(Y)\rangle$ to first non-trivial order in the fluctuation field $\zeta^\prime$ and the counter-terms $\Sigma,\Pi$, finding
\begin{eqnarray}
  G(X)&=&G(X)-\int d^2Y d^2Z G(X-Y)G(-Z)\left[4 D(Y-Z)G(Y-Z)-\Pi(Y-Z)\right]\,,\nonumber\\
  D(X)&=&D(X)-\int d^2Y d^2Z G(X-Y)G(-Z)\left[2 G^2(Y-Z)-\Sigma(Y-Z)\right]\,,
\end{eqnarray}
where I dropped the argument $i \zeta_0$ in $G(X,i\zeta_0)$ for clarity of notation. Note that if one sets $\Sigma(X)=0$ and $\Pi(X)=\nu^2 \delta(X)$ one recovers the R1 level equation (\ref{eq:r1geq}) for $G(X,i\zeta_0)$. Using straight lines to depict propagators $G(X)$ and wiggly lines to denote $D(X)$, the above equations suggest the solutions
\begin{equation}
  \label{eq:r2pis}
  \Pi(X)=4 D(X)G(X)= 4\ \vcenter{\!\hbox{\includegraphics[width=0.12\linewidth]{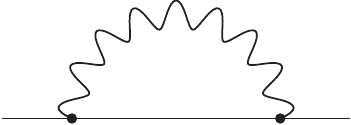}}}\,,\quad
  \Sigma(X)=2 G^2(X)= 2\ \vcenter{\!\hbox{\includegraphics[width=0.12\linewidth]{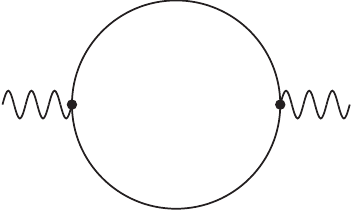}}}\,,
\end{equation}
where it should be noted that all propagators are themselves fully resummed, e.g. in momentum space
\begin{eqnarray}
  \label{eq:r2gs}
  \tilde G(K,i\zeta_0)&=&\int_X e^{-i K\cdot X} G(X,i\zeta_0)=\frac{1}{K^2+m^2+i \zeta_0+\tilde \Pi(K)}\,,\nonumber\\
  \tilde D(K)&=&\int_X e^{-i K\cdot X} D(X)=\frac{2\lambda}{1+2 \lambda \tilde \Sigma(K)}\,.
  \end{eqnarray}

The equations (\ref{eq:r2pis}), (\ref{eq:r2gs}) correspond to resummations of a class of diagrams beyond the mean-field resummation R1, hence they will be referred to as ``Resummation Level Two'' (``R2'') in the following. Note that in the $\tilde D(K)$ propagator for the $\zeta^\prime$ field, this reduces to the well-known RPA approximation, whereas in R2 this RPA approximation is coupled self-consistently with the approximation for the $\phi$ field. The partition function in the R2 approximation is obtained by including contributions to first non-trivial order in the fluctuation field $\zeta^\prime$ and the counter-terms $\Sigma,\Pi$, finding
\begin{eqnarray}
  Z_{R2}&=&\sqrt{\frac{\beta V}{16 \lambda \pi}}\int d\zeta_0 e^{-\beta V S^{\rm eff}_{R2}[i\zeta_0]}\,,\nonumber\\
  S^{\rm eff}_{R2}[i\zeta_0]&=&\frac{\zeta_0^2}{16 \lambda}+\frac{1}{2}\int_K\ln \left(K^2+m^2+i\zeta_0+\tilde \Pi(K)\right)-\frac{1}{2}\int_K \tilde \Pi(K)\tilde G(K)\nonumber\\
  &&+\frac{1}{2}\int_K\ln \left(1+2 \lambda \tilde \Sigma(K)\right)-\frac{1}{2}\int_K \tilde \Sigma(K)\tilde D(K)+\vcenter{\!\hbox{\includegraphics[width=0.12\linewidth]{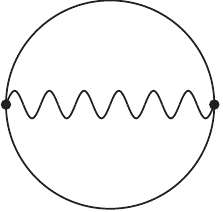}}}
%
%
\end{eqnarray}
As before, in the large volume limit the integral over $\zeta_0$ can be evaluated through its saddle at $i \zeta_0=z^*$ given by
\begin{equation}
  \label{eq:r2zeq}
  z^*_{R2}=4 \lambda G_{R2}(X=0,z^*_{R2})\,,
\end{equation}
cf. Eq.~(\ref{eq:r1zeq}). Note that contributions such as $\frac{\partial \tilde \Pi}{\partial i \zeta_0}$, $\frac{\partial \tilde \Sigma}{\partial i \zeta_0}$ have canceled out exactly in (\ref{eq:r2zeq}). Additionally, note that the two-loop contribution in $S_{R2}^{\rm eff}$ also cancels exactly,
\begin{equation}
  \label{eq:r2rel}
  -\frac{1}{2}\int_K \tilde \Sigma(K)\tilde D(K)+\vcenter{\!\hbox{\includegraphics[width=0.12\linewidth]{one-loopR2}}}=0\,,
\end{equation}
leaving only one-loop contributions to the R2 partition function.

Renormalization is again necessary to ensure a finite pole mass in $\tilde G(K,z^*_{R2})$ where it is advantageous to rewrite (\ref{eq:r2gs}) as
\begin{eqnarray}
  \label{eq:r2ginv}
  K^2+m^2+z^*_{R2}+\tilde \Pi(K)&=&K^2+m^2+3 z^*_{R2}-\delta \tilde \Pi(K)\,,\nonumber\\
  \delta \tilde \Pi(K)&=&8\lambda \int_Q \frac{2 \lambda \tilde \Sigma(Q)}{1+2 \lambda \tilde \Sigma(Q)} \tilde G(K-Q,z^*_{R2})\,,
\end{eqnarray}
such that the renormalization condition (\ref{eq:rencon}) is unchanged from the R1 level approximation.

\subsection{Resummation Level Three (``R3'')}

Since propagators in the action (\ref{eq:zgen}) have been fully resummed in R2, going beyond the R2 level approximation requires resummation of vertices. Taking the lessons learned from the R1 and R2 level approximations, the R2 scheme can be further refined by introducing vertex corrections $\delta \Gamma$ for the interaction term $\zeta^\prime \phi^2$ by rewriting (\ref{eq:zgen}) as
\begin{eqnarray}
  \label{eq:r3level}
  &S_0+S_I\rightarrow S_0^{\prime\prime} +S_I^{\prime\prime\prime}\,,&\\
  &S_I^{\prime\prime\prime}=-\frac{1}{2}\int d^2X d^2Y \left[\phi(X)\Pi(X-Y)\phi(Y)+\zeta^\prime(X)\Sigma(X-Y)\zeta^\prime(Y)\right]\,,\nonumber&\\
  &+i\int d^2X d^2Y d^2Z \zeta^\prime(X) \phi(Y)\phi(Z)\Gamma(X,Y,Z)-i\int d^2X d^2Y d^2Z \zeta^\prime(X) \phi(Y)\phi(Z)\delta \Gamma(X,Y,Z)&\,,\nonumber
\end{eqnarray}
where $\Gamma=1+\delta \Gamma$ and $S_0^{\prime\prime},G(X,i\zeta_0),D(X)$ are formally identical to the R2 expressions in Eqns.~(\ref{eq:r2level}) and $\Gamma=1+\delta \Gamma$. In order to fix the unknown vertex correction $\delta \Gamma(X,Y,Z)$, I calculate the connected three point function $\langle \zeta^\prime(X)\phi(Y)\phi(Z)\rangle$ to first non-trivial order in the fluctuation field $\zeta^\prime$ and counter-term $\delta \Gamma$. Expressing $\Gamma$ in Fourier-space through the two incoming momenta on the $\phi$ propagator lines as $\tilde \Gamma(P,K)$, this leads to
\begin{eqnarray}
  \label{eq:r3gamma}
  \delta \tilde \Gamma(P,K)=-4 &\int_Q& \tilde D(K+Q) \tilde G(Q)\tilde G(Q+P+K)\nonumber\\
  &&\times   \tilde \Gamma(Q,K)\tilde \Gamma(P,-P-K-Q)\tilde \Gamma(P+K+Q,-Q)\,,\nonumber\\
  \vcenter{\!\hbox{\includegraphics[width=0.12\linewidth]{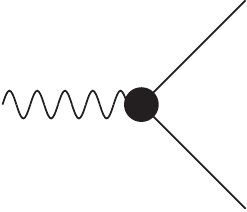}}}&=&1-4\vcenter{\!\hbox{\includegraphics[width=0.12\linewidth]{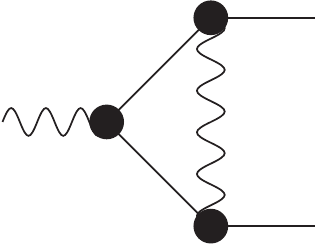}}}\,,
\end{eqnarray}
where a ``blob'' represents a fully resummed vertex. The unknown functions $\Pi,\Sigma$ are in turn fixed by calculating the two point functions $\langle \phi \phi\rangle$, $\langle \zeta^\prime \zeta^\prime\rangle$ to \textit{second} non-trivial order in the fluctuation field $\zeta^\prime$, second order in counter-terms $\Pi,\Sigma$ and first order in the counter-term $\delta \Gamma$. This leads to
\begin{equation}
  \label{eq:r3pis}
  \Pi(X)=4\ \vcenter{\!\hbox{\includegraphics[width=0.12\linewidth]{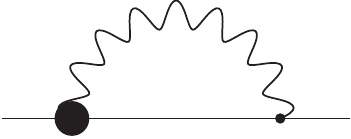}}}\,,\quad
  \Sigma(X)= 2\ \vcenter{\!\hbox{\includegraphics[width=0.12\linewidth]{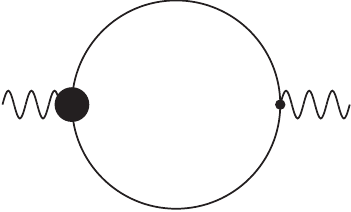}}}\,,
\end{equation}
where contributions have conspired in such a way that one vertex is fully resummed, whereas the other vertex is bare. The equations (\ref{eq:r2gs}), (\ref{eq:r3gamma}), (\ref{eq:r3pis}) correspond to the resummation of a class of diagrams beyond the R2 level approximation, hence they will be referred to as ``Resummation Level Three'' (``R3'') in the following. The partition function in the R3 approximation is obtained by including contributions to second non-trivial order in the fluctuation field $\zeta^\prime$ and counter-terms $\Pi,\Sigma$, and first order in the vertex counter-term $\delta \Gamma$, finding
\begin{eqnarray}
  \label{eq:zr3}
  Z_{R3}&=&\sqrt{\frac{\beta V}{16 \lambda \pi}}\int d\zeta_0 e^{-\beta V S^{\rm eff}_{R3}[i\zeta_0]}\,,\nonumber\\
  S^{\rm eff}_{R3}[i\zeta_0]&=&\frac{\zeta_0^2}{16 \lambda}+\frac{1}{2}\int_K\ln \left(K^2+m^2+i\zeta_0+\tilde \Pi(K)\right)-\frac{1}{2}\int_K \tilde \Pi(K)\tilde G(K)\nonumber\\
  &&+\frac{1}{2}\int_K\ln \left(1+2 \lambda \tilde \Sigma(K)\right)
  +2\ \vcenter{\!\hbox{\includegraphics[width=0.12\linewidth]{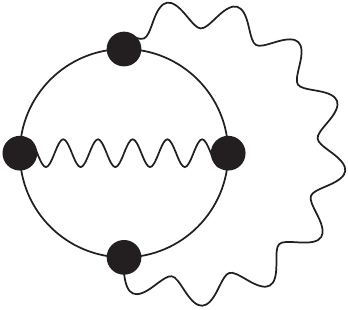}}}
\end{eqnarray}
where I used the R3 equivalent of (\ref{eq:r2rel}) to rewrite $S_{R3}^{\rm eff}$. Using the form of the resummed vertex (\ref{eq:r3gamma}) it is possible to rewrite the formally three-loop contribution to the partition function as a one-loop integral as
\begin{eqnarray}
  \label{eq:deltaS}
  2\ \vcenter{\!\hbox{\includegraphics[width=0.12\linewidth]{one-loopR3}}} &=& -\frac{1}{4}\int_K \tilde D(K)  \delta\tilde\Sigma(K)\,,\nonumber\\
   \delta\tilde \Sigma(K) &=& 2\ \vcenter{\!\hbox{\includegraphics[width=0.12\linewidth]{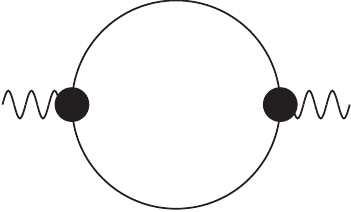}}}-2 \ \vcenter{\!\hbox{\includegraphics[width=0.12\linewidth]{one-loopSigmar3}}}\,.
  \end{eqnarray}

The large volume limit of the partition function is again determined by the saddle point $i\zeta_0=z^*$ given by
\begin{equation}
  \label{eq:r3zeq}
  z_{R3}^*=4 \lambda G_{R3}(X=0,z_{R3}^*)\,,
\end{equation}
cf. Eqns.~(\ref{eq:r1zeq}), (\ref{eq:r3zeq}). Note that contributions such as $\frac{\partial \tilde \Pi}{\partial i \zeta_0}$, $\frac{\partial \tilde \Sigma}{\partial i \zeta_0}$ and $\frac{\partial \tilde \Gamma}{\partial i \zeta_0}$ have canceled out exactly in Eq.~(\ref{eq:r3zeq}).  Renormalization proceeds as for R2 with Eq.~(\ref{eq:r2gs}) rewritten as
\begin{eqnarray}
  \label{eq:r3inv}
  K^2+m^2+z^*_{R2}+\tilde \Pi(K)&=&K^2+m^2+3 z^*_{R2}-\delta \tilde \Pi(K)\,,\\
  \delta \tilde \Pi(K)&=&8\lambda \int_Q \left[\frac{2 \lambda \tilde \Sigma(Q)}{1+2 \lambda \tilde \Sigma(Q)}\tilde \Gamma(K,Q-K)-\delta \Gamma(K,Q-K)\right]\tilde G(K-Q,z^*_{R2})\,,\nonumber
\end{eqnarray}
such that the renormalization condition (\ref{eq:rencon}) is unchanged from the R1 and R2 level approximation.

\subsection{Beyond R3}

How to go beyond the R3-level approximation scheme? On the one hand, it is possible to improve upon the results found in the R0-R3 schemes by performing a power series expansion of $S_I$. Taking the example of R0, this would imply performing an expansion, not around free field theory, but around a field theory with a non-perturbative mass parameter $i\zeta_0$. This approach will require the calculation of diagrams beyond one-loop order and I suspect that it will run into the same kind of difficulties that beset ordinary perturbation theory, such as divergent series character, results oscillating with truncation order of the series, and so on.

On the other hand, it is conceivable that the present schemes may be generalized by noting that the type of diagrams not fully included in the R3-level approximations are those that correspond to resummed 4-vertices. This suggests that a potential R4-level approximation scheme may be constructed by adding and subtracting an effective 4-vertex $\Gamma^{(4)}$ to the  action (\ref{eq:zgen}) and calculating the connected 4-point functions to first non-trivial order in the auxiliary field and counter-terms. A similar procedure would then be employed for R5 (5-point functions), R6 (6-point functions) and so on. Progressing in this fashion,  it is conceivable that one could generalize the R-level schemes to higher accuracy and retain attractive features such as requiring only the calculation of one-loop integrals.

\section{Results}

The pole mass $M$ (defined by the pole of the two-point function $\left(\langle\phi\phi\rangle(K^2=-M^2)\right)^{-1}=0$) is amenable to precision calculation in 1+1 $\phi^4$ theory and hence is an interesting quantity to test the present resummation scheme. After renormalization, the pole mass can be expressed as a function of the dimensionless coupling
$$g\equiv \frac{\lambda}{m_R^2}\,,$$
and the theory becomes gap-less at a critical coupling $g_c$ with critical exponent $c_1$ as
$$
M(g)\propto |g-g_c|^{c_1}\,,\quad g\rightarrow g_c\,.
$$
Similarly, at the critical coupling the two-point function scales as
$$
\langle \phi(x)\phi(0)\rangle\propto \left(\frac{1}{x^2}\right)^{c_2}\,,\quad g\rightarrow g_c\,,
$$
where the exact results are $c_1=1,c_2=\frac{1}{8}$ \cite{PhysRev.65.117}. Another interesting observable is the vacuum energy, defined as
\begin{equation}
  \Lambda=-\frac{1}{\beta V}\left(\ln Z(g)-\ln Z(g=0)\right)
\end{equation}
Using the scheme choice $\bar\mu^2=m_R^2$, the pole mass in the R0, R1 approximations simply becomes $M^2=m_R^2$, hence it is not interesting to discuss these approximation schemes here.

\begin{figure}[t]
  \includegraphics[width=\linewidth]{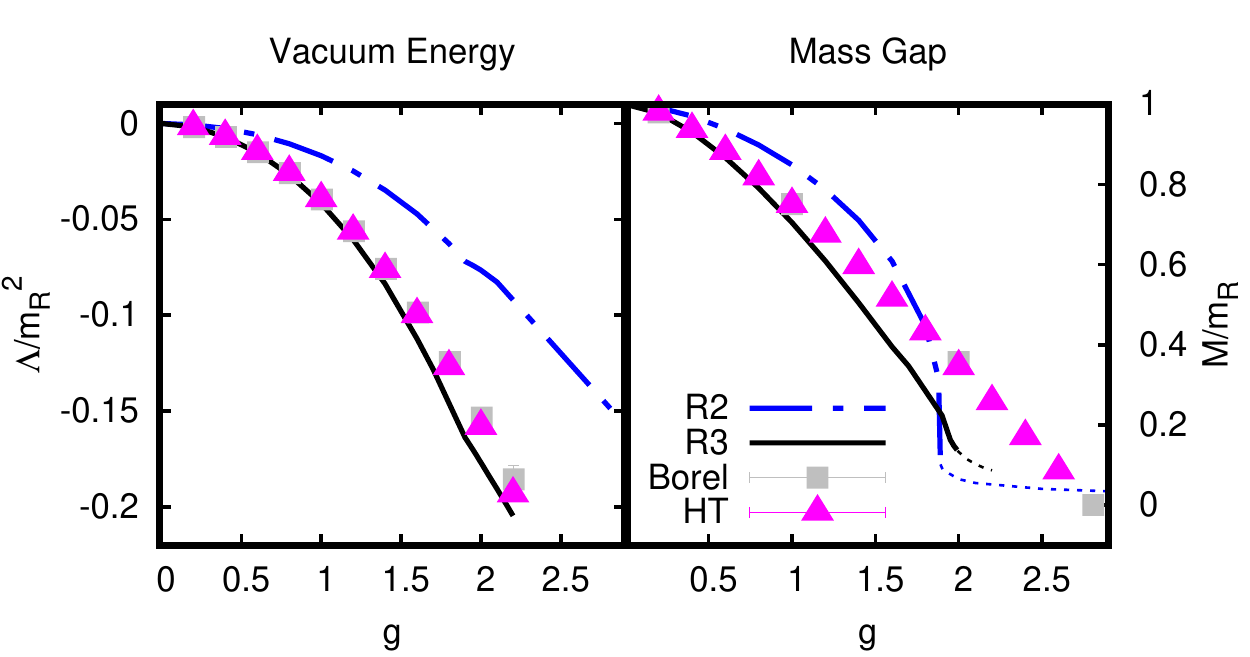}
  \caption{\label{fig:one} Vacuum energy  and mass gap for scalar $\phi^4$ in 1+1 dimensions as a function of dimensionless coupling $g=\frac{\lambda}{m_R^2}$. Results are shown for the R2 and R3 approximation (dashed and full lines, respectively) and precision calculations from Borel summation ('Borel') and Hamiltonian truncation ('HT') from Refs.~\cite{Serone:2018gjo,Elias-Miro:2017xxf}, respectively. Mass-gap results for R2 and R3 results become resolution-dependent for small mass (dotted lines). See text for details.} 
\end{figure}

For the R2 scheme, the pole mass $M$ is formally defined through the inverse two-point function in Fourier space (\ref{eq:r2ginv}) as
\begin{equation}
\label{eq:dmf}
  M^2=m_R^2+\delta m^2-\delta\tilde \Pi(K^2=-M^2)\,,\quad \delta m^2=12 \lambda\int_K
  \left[\tilde G(K,z_{R2}^*)-\frac{1}{K^2+m_R^2}\right]\,.
\end{equation}
The R2 scheme is amenable to efficient numerical evaluation using Gauss-Legendre integration (details for the numerical evaluation can be found in appendix \ref{sec:appr2}). In this scheme, values for $\delta \Pi$ are calculated and stored for positive $K^2$, and linear extrapolation is used to obtain $\delta\tilde \Pi(K^2=-M^2)$. Results for $M$ and $\Lambda$ as a function of $g$ are shown in Fig.~\ref{fig:one}, where also comparison to recent precision calculations of the same quantities from Ref.~\cite{Elias-Miro:2017xxf,Serone:2018gjo} are shown.

As can be seen from Fig.~\ref{fig:one}, R2 results for the mass gap and vacuum energy are qualitatively similar to those obtained from Refs.~\cite{Elias-Miro:2017xxf,Serone:2018gjo}, but show clear quantitative deviations. Moreover, the R2 scheme predicts a phase transition close to $g_c\simeq 1.89$ with a critical exponent $c_1$ that is not consistent with the exact value $c_1=1$. (Curiously, the R2 value $c_2= 0.12(5)$ obtained numerically at $g_c\simeq 1.89$ seems to be close to the exact value). Similar behavior has been reported in the RPA approximation in Ref.~\cite{Hansen:2002zt}, and thus the behavior for R2 seen in Fig.~\ref{fig:one} is thus not too surprising given that the R2 approximation employs an RPA approximation in the $\zeta^\prime$ propagator.

The R3 approximation is considerably closer to the results from precision calculations for the vacuum energy and mass gap, and indeed seems to provide a quantitatively reliable approximation at intermediate coupling. However, there are marked quantitative differences between the precision calculations and the R3 approximation, notable concerning the critical coupling $g_c$. While recent works such as Refs.~\cite{Schaich:2009jk,Bosetti:2015lsa,Elias-Miro:2017xxf,Serone:2018gjo} consistently find $g_c\simeq 2.8$, the critical coupling in the R3 approximation is approximately $g_c\simeq 2.2$. Curiously, both the vacuum energy and the critical exponent $c_2=0.12(5)$ in the R3 approximation are numerically close to the corresponding results from precision calculations.

For convenience, the source for the R2 and R3 level algorithms described here is publicly available at \cite{codedown}.

\section{Summary and Conclusions}

In this work, I have presented a series of non-perturbative approximation schemes. These schemes are simple to implement, because they only require evaluation of one-loop integrals.

To test the quality of these approximation schemes, I considered scalar $\phi^4$ theory in 1+1 dimensions at zero temperature, finding approximate agreement for the vacuum energy and mass gap with precision results from other methods at intermediate values of the interaction. Since the well-known mean-field resummation scheme utterly fails at reproducing these observables, it seems that the resummation schemes R2 and R3 presented here encode information about the quantum field theory at intermediate interaction strength \textit{approximately}.

Because the resummation schemes R0-R3 are all approximate in nature, they will not be useful for calculating quantities where other, in particular \textit{exact}, methods are available, such as in the calculation of critical exponents in $\phi^4$ theory \cite{Cappelli:2018vir}. However, R0-R3 may serve as a general purpose tool for calculating properties of quantum field theories \textit{approximately} at intermediate interaction strength that are inaccessible by other methods. In particular, this could be the case for calculating properties of scalar $\phi^4$ in 2+1 and 3+1 dimension, finite-temperature properties (including transport), or properties for theories with similar Lagrangians, such as the Gross-Neveu model \cite{Gross:1974jv} or maybe even gauge theories along the lines of \cite{Akerlund:2016his}.

\section{Acknowledgments}

This work was supported in part by the Department of Energy, DOE award No DE-SC0017905. I would like to thank Marcus Pinto for helpful discussions and urging me to write this up note and Philipp Stanzer and Mark Watson for spotting typos. I also would like to thank Joan Elias Miro and Gabriele Spada for providing the data points from Refs.~\cite{Elias-Miro:2017xxf,Serone:2018gjo}, respectively.

\begin{appendix}
  \section{Details of the Numerical Algorithm for the R2 and R3 schemes}
  \label{sec:appr2}

  In this appendix, details on the numerical evaluation of the R2 and R3 approximation schemes are given. For notational brevity, I choose to describe the R3 scheme and note which steps to omit in order to get the R2 approximation. 

  As pointed out in the main text, the numerical schemes require the evaluation of one-loop integrals, e.g. of the form
  \begin{eqnarray}
    \label{eq:appi1}
    \tilde \Sigma(K)&=&2\int \frac{d^2 q}{(2\pi)^2} \tilde G(K-Q)\tilde G(K) \tilde \Gamma(Q,K-Q)\,,\nonumber\\
    \delta \Pi(K)&=&8 \lambda \int \frac{d^2 q}{(2\pi)^2} \left[\tilde G(Q) \tilde T(K-Q) \tilde \Gamma(K,-Q)-\tilde G(Q)\delta \Gamma(K,-Q)      \right]\,,\nonumber\\
    \delta\tilde\Gamma(P,K)&=&-8\lambda \int \frac{d^2 q}{(2\pi)^2}\frac{1}{1+2\lambda \tilde \Sigma(K+Q)}\tilde G(Q)\tilde G(Q+P+K)\nonumber\\
    &&\times   \tilde \Gamma(Q,K)\tilde \Gamma(P,-P-K-Q)\tilde \Gamma(P+K+Q,-Q)\,,
  \end{eqnarray}
cf. Eqns.~(\ref{eq:r3gamma}),(\ref{eq:r3inv}) where
  \begin{equation}
    \tilde G(K)=\frac{1}{K^2+m^2+3 z_R^*-\delta \Pi(K)}\,,\quad
    \tilde T(K)=\frac{2 \lambda \tilde \Sigma(K)}{1+2\lambda \tilde \Sigma(K)}\,,
  \end{equation}
  and where $\Gamma(Q,P)=1$  ($\delta \Gamma(Q,P)=0$) in the R2 scheme. 
  Renormalization is explicitly performed by replacing
  \begin{equation}
    \label{eq:deltam}
  K^2+m^2+3 z_R^*=K^2+m_R^2+\delta m^2\,,\quad
  \delta m^2=12 \lambda \int \frac{d^2 q}{(2 \pi)^2}\left(\tilde G(Q)-\frac{1}{Q^2+m_R^2}\right)\,,
  \end{equation}
  cf. Eq.~(\ref{eq:dmf}).
Using mass units $m_R=1$,  I find it advantageous to compactify the integration domain by using $q_0=\tan\left(\frac{\pi x}{2}\right)$ such that for instance
\begin{equation}
  \label{eq:intex1}
  \delta m^2=\frac{12 \lambda}{16} \int_{-1}^1 dx_0 \int_{-1}^1 dx_1 \frac{1}{\cos^2\left(\frac{\pi x_0}{2}\right)\cos^2\left(\frac{\pi x_1}{2}\right)}\left(\tilde G({\bf x}^2)-\frac{1}{{\bf x}^2+1}\right)\,,
  \end{equation}
  where ${\bf x}^2\equiv \tan^2\left(\frac{\pi x_0}{2}\right)+\tan^2\left(\frac{\pi x_1}{2}\right)$. Integrals such as this one are evaluated 
numerically as quadratures with $N$ stencils $x_1,x_2,\ldots,x_{N}$ and weights $w_1,w_2,\ldots w_{N}$. I find it convenient to use Gauss-Legendre quadrature where the stencils are given by the roots of the Legendre polynomial of order $N$ and standard weights
\begin{equation}
P_{N}(x_i)=0\,,\quad  w_i=\frac{2}{(1-x_i^2)\left(P_{N}^\prime(x_i)\right)^2}\,.
\end{equation}
I find that tabulated values for $x_i$ can be efficiently obtained to high precision for $N$ up to $N\simeq 2000$. In practice, I find that $N\simeq 200$ is typically sufficient for obtaining results of percentage precision, and I chose $N$ odd in order to include the point $x_i=0$ in the set of stencils. To be explicit, for the results shown in Fig.~\ref{fig:one}, I used $N=501$ for R2 and $N=201$ for R3.

Once the number of stencils has been chosen, integrals such as (\ref{eq:intex1}) become
$$
\delta m^2=\frac{12 \lambda}{16} \sum_{i,j=1}^N \tilde w_i \tilde w_j \left(\tilde G(x_{ij})-\frac{1}{x_{ij}+1}\right)\,,
$$
where $x_{ij}\equiv \tan^2\left(\frac{\pi x_i}{2}\right)+\tan^2\left(\frac{\pi x_j}{2}\right)$ and I used modified weights
$$
\tilde w_i=\frac{w_i}{\cos^2\left(\frac{\pi x_i}{2}\right)}\,.
$$
Evaluation of $\tilde G(x_{ij})$ (and similarly $\tilde T(x_{ij})$ is done by discretizing the self-energies $\delta \Pi(K)$, $\Sigma(K)$, which at zero temperature only depend on $K^2$. While it would be possible to re-use the discretization $K^2\rightarrow K^2_i=\tan\left(\frac{\pi x_i}{2}\right)$ from above, I find it advantageous to only store self-energies up to a certain maximal value of $K^2_{\rm max}$, because for a fixed number of Legendre stencils, numerical errors quickly accumulate for $K^2\gtrsim \frac{3}{4} \tan\left(\frac{\pi x_N}{2}\right)$. Hence I use the discretization $\delta \Pi(K^2)\rightarrow \delta \Pi_i=\delta \Pi\left(\tan\left(x_i \arctan (K_{\rm max}^2)\right)\right)$. In practice, I find that for $K^2_{\rm max}>1000 m_R^2$, results no longer depend on the specific value of $K^2_{\rm max}$.
For the vertices (not needed in R2), I use a discretization that stores $\tilde \Gamma(P,K)$ as a function of $P_i^2=\tan\left(\frac{\pi x_i}{2}\right)$, $K_i^2=\tan\left(\frac{\pi x_i}{2}\right)$ and the relative angle $\phi$ between these two vectors, $\cos\phi=\frac{P\cdot K}{\sqrt{P^2 K^2}}$. For the angular discretization, I use Fourier-stencils $\phi\rightarrow \phi_l=\frac{l \pi}{N_\phi}$ with $l=0,1,\ldots 2 N_\phi-1$. In practice, I find that $N_\phi=48$ leads to acceptable accuracy in the numerical evaluations, and this choice was also used for the R3 scheme in Fig.~\ref{fig:one}.

Integrals such as (\ref{eq:appi1}) also require evaluation of self-energies and vertices in-between stencils, for which I use linear interpolation. Also, results for $K^2>K^2_{\rm max}$ are needed for which I set $\delta \Pi,\delta \Gamma$ to zero, while for $\Sigma$ I use the analytic result when neglecting the internal self-energy and vertex correction
\begin{equation}
  \Sigma(K^2>K^2_{\rm max})\simeq \frac{2}{\pi} \frac{{\rm arctanh}{\sqrt{\frac{K^2}{K^2+4 m_R^2+4 \delta m^2}}}}{K^2}\,.
\end{equation}

The algorithm to obtain observables in the R3 (R2) scheme is then as follows:
\begin{enumerate}
\item
  Load quadrature data for $N$ stencils $x_i$ and weights $w_i$
\item
  Calculate modified stencils such as $\tan\left(\frac{\pi x_i}{2}\right)$ and weights $\frac{w_i}{\cos^2\left(\frac{\pi x_i}{2}\right)}$
\item
  Set initial values for $\delta \Pi,\Sigma,\delta \Gamma,\delta m^2$ to zero
\item
  Using the known values for self-energies and vertices, calculate new $\Sigma(K^2)$ from (\ref{eq:appi1})
\item
 R3 only: using the known value for self-energies and vertices, calculate new $\delta \Gamma$ from (\ref{eq:appi1}). For the stability of the algorithm, it is advantageous to replace $\delta \Gamma$ by the sum of half of its old and new values, respectively
\item
  Using the known values for self-energies and vertices, calculate new $\delta \Pi(K^2)$ from (\ref{eq:appi1})
\item
  Using the known values for self-energies and vertices, calculate the new value for $\delta m^2$ by re-writing (\ref{eq:deltam}) as
  $$
  \delta m^2=\frac{12 \lambda \int_K \frac{\delta \tilde \Pi(K^2)}{K^2+1} \tilde G(K)}{1+12 \lambda \int_K \frac{\tilde G(K)}{K^2+1}}
  $$
  and using root bracketing for $\delta m^2\in [{\rm max}(\delta \tilde \Pi(0)-m_R^2,0),\delta m^2_{\rm max}]$ to solve this equation. In practice, choosing $\delta m^2_{\rm max}<10 m_R^2$ turns out to be sufficient for most purposes.
\item
  Repeat from step 4 until converged (will take longer for larger interaction strength)
\item
  Using the converged self-energies, vertices and $\delta m^2$, calculate observables (see below)
  \end{enumerate}

With self-energies and mass shift $\delta m^2$ converged, one can obtain an estimate for the pole mass $M$ by linear extrapolation of $\delta \tilde \Pi$, giving
\begin{equation}
  M\simeq \sqrt{\frac{1+\delta m^2-\delta \tilde \Pi(0)}{1-\frac{\delta \tilde\Pi_1-\delta \tilde \Pi_0}{\tan(x_1\arctan(K^2_{\rm max}))}}}\,,
  \end{equation}
cf. ~\ref{eq:dmf}, where $\delta \tilde \Pi_1$ is the value of the self-energy correction at the first non-vanishing stencil value $\tan(x_1\arctan(K^2_{\rm max}))$. Results for $M$ obtained in this fashion are shown in Fig.~\ref{fig:one}.

For the vacuum energy, it is advantageous to rewrite the expression (\ref{eq:zr3}) by splitting
\begin{eqnarray}
\frac{1}{2} \int_K \tilde \Pi(K)\tilde G(K)&=&\frac{1}{4}\int_K \tilde \Pi(K)\tilde G(K)+\frac{1}{2}\int_K \tilde \Sigma(K)\tilde D(K)\nonumber\\
  &=&\frac{z^{*2}}{8\lambda}-\frac{1}{4}\int_K \delta \tilde \Pi(K)\tilde G(K)+\lambda\int_K \frac{\tilde \Sigma(K)}{1+2\lambda \tilde \Sigma(K)}
\end{eqnarray}
such that (\ref{eq:zr3}) becomes
\begin{eqnarray}
  S_{R3}^{\rm eff}[z^*]&=&C_\Pi+C_V+C_\Sigma+C_\Sigma^\prime\,,\nonumber\\
  C_\Pi&=&\frac{1}{2}\int \frac{d^2k}{(2 \pi)^2}\left[\ln \frac{K^2+m_R^2+\delta m^2-\delta \tilde \Pi(K)}{K^2+m_R^2+\delta m^2}+\frac{1}{2}\delta \tilde \Pi(K)\tilde G(K)\right]\,,\nonumber\\
  C_V&=&-\frac{3z^{* 2}}{16 \lambda}+\frac{1}{2}\int_K \ln \left(K^2+m_R^2+\delta m^2\right)\,,\nonumber\\
  C_\Sigma&=&\frac{1}{2}\int \frac{d^2k}{(2\pi)^2}\left[\ln\left(1+2 \lambda \tilde \Sigma(K)\right)-\frac{2 \lambda \tilde \Sigma(K)}{1+2 \lambda \tilde \Sigma(K)} \right]\,,\nonumber\\
  C_\Sigma^\prime&=&-\frac{1}{4}\int \frac{d^2k}{(2\pi)^2}\frac{\delta \tilde \Sigma}{1+2\lambda \tilde \Sigma(K)}\,,
\end{eqnarray}
where the modified self-energy $\delta \tilde \Sigma(K)$ from (\ref{eq:deltaS}) is
$$
\delta \tilde \Sigma(K)=2 \int \frac{d^2q}{(2\pi)^2}\tilde G(K-Q)\tilde(K)\tilde \Gamma(Q,K-Q)\left(\tilde\Gamma(Q,K-Q)-1\right)\,.
$$
Note that $\delta \Sigma$ vanishes for R2 as does the contribution $C_\Sigma^\prime$. Only the contribution $C_V$ contains divergencies, while all the other ones are finite and are calculated using the quadrature scheme discussed above. Using $m^2+3 z^*=m_R^2+\delta m^2$ from (\ref{eq:deltam}) and the renormalization condition (\ref{eq:rencon}), $C_V$ can be calculated as
\begin{eqnarray}
  C_V&=&-\frac{(3 z^{*})^2}{48 \lambda}+J_0\left(\sqrt{m_R^2+\delta m^2}\right)\,,\nonumber\\
  &=&-\frac{m^4}{48\lambda}-\frac{(\delta m^2)^2}{48 \lambda}+\frac{m_R^4}{48\lambda}+\frac{m_R^2+\delta m^2}{8\pi} \ln \frac{m_R^2 e^1}{m_R^2+\delta m^2}\,,
\end{eqnarray}
Renormalizing this contribution by demanding that $C_V(\lambda=0)=0$, this leads to
$$
C_V^{\rm ren}=-\frac{(\delta m^2)^2}{48 \lambda}+\frac{m_R^2+\delta m^2}{8\pi} \ln \frac{m_R^2 e^1}{m_R^2+\delta m^2}-\frac{m_R^2}{8\pi}\,.
$$
Summing up the contributions $C_V^{\rm ren},C_\Pi,C_\Sigma,C_\Sigma^\prime$ leads to the results for the vacuum energy shown in Fig.~\ref{fig:one}.

For convenience, the source for the R2 and R3 level algorithms described here is publicly available at \cite{codedown}.

  \end{appendix}

\bibliographystyle{JHEP}
\bibliography{lambda}

\providecommand{\href}[2]{#2}\begingroup\raggedright\begin{thebibliography}{10}

\bibitem{Laine:2016hma}
M.~Laine and A.~Vuorinen, {\it {Basics of Thermal Field Theory}},  {\em Lect.
  Notes Phys.} {\bf 925} (2016) pp.1--281,
  [\href{http://arxiv.org/abs/1701.01554}{{\tt arXiv:1701.01554}}].

\bibitem{Carmi:2018qzm}
D.~Carmi, L.~Di~Pietro, and S.~Komatsu, {\it {A Study of Quantum Field Theories
  in AdS at Finite Coupling}},  \href{http://arxiv.org/abs/1810.04185}{{\tt
  arXiv:1810.04185}}.

\bibitem{Loinaz:1997az}
W.~Loinaz and R.~S. Willey, {\it {Monte Carlo simulation calculation of
  critical coupling constant for continuum phi**4 in two-dimensions}},  {\em
  Phys. Rev.} {\bf D58} (1998) 076003,
  [\href{http://arxiv.org/abs/hep-lat/9712008}{{\tt hep-lat/9712008}}].

\bibitem{Schaich:2009jk}
D.~Schaich and W.~Loinaz, {\it {An Improved lattice measurement of the critical
  coupling in phi(2)**4 theory}},  {\em Phys. Rev.} {\bf D79} (2009) 056008,
  [\href{http://arxiv.org/abs/0902.0045}{{\tt arXiv:0902.0045}}].

\bibitem{Bosetti:2015lsa}
P.~Bosetti, B.~De~Palma, and M.~Guagnelli, {\it {Monte Carlo determination of
  the critical coupling in $\phi^4_2$ theory}},  {\em Phys. Rev.} {\bf D92}
  (2015), no.~3 034509, [\href{http://arxiv.org/abs/1506.08587}{{\tt
  arXiv:1506.08587}}].

\bibitem{PhysRev.65.117}
L.~Onsager, {\it Crystal statistics. i. a two-dimensional model with an
  order-disorder transition},  {\em Phys. Rev.} {\bf 65} (Feb, 1944) 117--149.

\bibitem{Elias-Miro:2017xxf}
J.~Elias-Miro, S.~Rychkov, and L.~G. Vitale, {\it {High-Precision Calculations
  in Strongly Coupled Quantum Field Theory with Next-to-Leading-Order
  Renormalized Hamiltonian Truncation}},  {\em JHEP} {\bf 10} (2017) 213,
  [\href{http://arxiv.org/abs/1706.06121}{{\tt arXiv:1706.06121}}].

\bibitem{Serone:2018gjo}
M.~Serone, G.~Spada, and G.~Villadoro, {\it {$\lambda \phi^4$ Theory I: The
  Symmetric Phase Beyond NNNNNNNNLO}},  {\em JHEP} {\bf 08} (2018) 148,
  [\href{http://arxiv.org/abs/1805.05882}{{\tt arXiv:1805.05882}}].

\bibitem{Hansen:2002zt}
H.~Hansen, G.~Chanfray, D.~Davesne, and P.~Schuck, {\it {Random phase
  approximation and extensions applied to a bosonic field theory}},  {\em Eur.
  Phys. J.} {\bf A14} (2002) 397--411,
  [\href{http://arxiv.org/abs/hep-ph/0201279}{{\tt hep-ph/0201279}}].

\bibitem{codedown}
P.~Romatschke, {\it {Numerical codes for $\phi^4$ theory in 1+1 dimensions at
  R2/R3 level}},  {\em https://github.com/paro8929/Resummation$\quad$}.

\bibitem{Cappelli:2018vir}
A.~Cappelli, L.~Maffi, and S.~Okuda, {\it {Critical Ising Model in Varying
  Dimension by Conformal Bootstrap}},  {\em JHEP} {\bf 01} (2019) 161,
  [\href{http://arxiv.org/abs/1811.07751}{{\tt arXiv:1811.07751}}].

\bibitem{Gross:1974jv}
D.~J. Gross and A.~Neveu, {\it {Dynamical Symmetry Breaking in Asymptotically
  Free Field Theories}},  {\em Phys. Rev.} {\bf D10} (1974) 3235.

\bibitem{Akerlund:2016his}
O.~Akerlund and P.~de~Forcrand, {\it {Mean distribution approach to spin and
  gauge theories}},  {\em Nucl. Phys.} {\bf B905} (2016) 1--15,
  [\href{http://arxiv.org/abs/1601.01175}{{\tt arXiv:1601.01175}}].

\end{thebibliography}\endgroup
\end{document}